\documentclass[12pt]{article}
\usepackage{color}
\usepackage{amsmath}
\usepackage{cite}
\usepackage[pdftex]{graphicx}
\begin{document}    
\vspace*{1cm}

\renewcommand\thefootnote{\fnsymbol{footnote}}
\begin{center} 
  {\Large\bf Neutrino models with a zero mass eigenvalue}
\vspace*{1cm}

{\Large Daijiro Suematsu}\footnote[1] {professor emeritus, ~e-mail:
suematsu@hep.s.kanazawa-u.ac.jp}
\vspace*{0.5cm}\\

{\it Institute for Theoretical Physics, Kanazawa University, 
Kanazawa 920-1192, Japan}
\end{center}
\vspace*{1.5cm} 

\noindent
{\Large\bf Abstract}\\
Absolute values of the neutrino mass are not known still now although their upper bounds are
constrained through several experiments and observations.
Recent analyses of cosmological observations present severe constraint on the sum of neutrino 
masses. It might suggest an interesting possibility for the absolute 
values of neutrino mass and their ordering. In this paper, taking it as a useful hint, 
we study possible neutrino models with a zero mass eigenvalue from a view point of neutrino oscillation 
data and baryon number asymmetry in the Universe. We focus our study on the seesaw type mass 
generation by making a certain assumption for origin of right-handed neutrino mass.

\newpage
\setcounter{footnote}{0}
\renewcommand\thefootnote{\alph{footnote}}

\section{Introduction}
The existence of nonzero neutrino mass has been confirmed through various neutrino 
oscillation experiments.
Neutrino oscillations constrain squared mass differences and flavor mixing \cite{pdg}.
Although they cannot fix their absolute values, analyses based on the 
measurements of tritium $\beta$ decay \cite{katrin}, neutrinoless double $\beta$ 
decay \cite{kamzen}, and also cosmological observations \cite{cmb} give us useful 
informations for their upper bounds. 
Especially, recent cosmological data derived from analysis of the cosmic microwave background 
(CMB) and the baryon acoustic oscillations (BAO) impose stringent upper bound on the sum 
of neutrino masses. 
After freezeout of the weak interaction, neutrinos fill the Universe as the cosmic 
background which can present a crucial prediction for the cosmological model \cite{cneurev}.
Free streaming length of this cosmic neutrinos which is determined by the neutrino mass 
affects the scale of large scale structure. Cosmic neutrino energy density fixed by the neutrino 
mass contributes to the expansion rate of the Universe and also photons are affected by the cosmic 
neutrino background through their travel from the last scattering surface to today.
Since cosmological observables relevant to these are connected to the neutrino mass, 
stringent constraint on the sum of the neutrino mass can be derived through the observation of them.

Upper bound presented for it by recent DESI result \cite{desi} is 
very impressive even though it depends on the assumed cosmological model and the analysis. 
It suggests that the inverted mass ordering (IO) is ruled out and the lightest neutrino mass 
is favored to be zero in the normal mass ordering (NO).
Subsequent studies \cite{quel} suggest that the stringent bound in the DESI analysis
is caused by the use of Planck data including lensing anomalies and the BAO 
data at $z=0.7$ and the bound becomes weaker if these are removed from the
analysis and replaced by the new one. 
It is also pointed out that assuming the dynamical dark energy model instead of 
$\Lambda$CDM and using the hierarchical neutrino mass instead of the averaged neutrino mass  
can make the bound weaker.
New data expected to be obtained in a few years will improve the statistics and these 
uncertainties.
Although it may be considered to be premature to discuss neutrino phenomenology 
based on it, it could motivate us to consider neutrino models with a zero mass 
eigenvalue as a promising possibility and to discuss flavor structure of neutrino Yukawa 
couplings using such models.
 
In the generation of Majorana neutrino mass based on the seesaw mechanism \cite{seesaw},
it is well-known that a mass eigenvalue is zero if only two right-handed neutrinos are
introduced to the SM. In the present study, we fix origin of the mass of right-handed neutrinos 
and extend the model under an assumption of tribimaximal neutrino mixing 
to the ones with three right-handed neutrinos keeping a zero mass eigenvalue. 
This extension is found to be supported by imposing that the baryon number asymmetry is 
generated through thermal leptogenesis due to decay of the lightest right-handed neutrino.
We discuss features of leptogenesis focusing on thermal generation of the lightest right-handed 
neutrino and washout of generated lepton number asymmetry.
 
Remaining parts of the paper are organized as follows.
In section 2 we present possible neutrino models with three right-handed neutrinos, 
which can realize both a zero mass eigenvalue and flavor mixing favored by the neutrino 
oscillation data. In section 3 we discuss constraints on neutrino Yukawa couplings based 
on the neutrino oscillation data and the generation of baryon number asymmetry through 
leptogenesis. We summarize the paper in section 4.      
 
\section{Neutrino models with a zero mass eigenvalue}
We consider an extension of the SM by adding Yukawa interaction terms for neutrinos such as
\begin{eqnarray}
{\cal L}_y&\supset& \sum_k\left[\sum_{\alpha=e,\mu.\tau}
h_{\alpha k}^\nu\bar\ell_\alpha \varphi N_k
+y_kS\bar N_kN_k^c  + {\rm h.c.}\right],
\label{yukawa}
\end{eqnarray}
where $\ell_\alpha$ and $N_k$ are a left-handed doublet lepton and a singlet 
right-handed neutrino. 
A doublet scalar and a singlet scalar of SU(2) are represented by $\varphi$ and $S$, 
respectively.
Right-handed neutrino mass is supposed to be generated through a second term 
as $M_{N_k}=y_k u$ where $u$ is an absolute value of the vacuum expectation value (VEV) of $S$
and defined as $\langle S\rangle=ue^{i\gamma}$.
Yukawa coupling constants $h^\nu_{\alpha k}$ and $y_k$ are assumed to be real, and $y_k$ 
realizes mass ordering $M_{N_k}<M_{N_{k+1}}$.
Since $CP$ symmetry is assumed to be broken 
through the VEV of $S$, potential of $S$ violates lepton number so that the 
appearance of a massless Majoron is evaded. 
 Light mass of neutrinos derived from 
these Yukawa couplings in eq.~(\ref{yukawa}) is generated through a so-called seesaw 
mechanism \cite{seesaw} and has Majorana nature. 
A formula of its mass matrix can be generally expressed as
\begin{equation}
({\cal M}_\nu)_{\alpha\beta}=\sum_k h_{\alpha k}^\nu h_{\beta k}^\nu e^{i\gamma}\Lambda_k,
\end{equation}
where $\Lambda_k$ is a relevant mass scale which depends on the mass generation scheme, and
a common complex phase $\gamma$ is caused. 
Here, as its representative examples, we consider the expression for $\Lambda_k$ in two models, 
that is, the type-I seesaw model \cite{seesaw} and the scotgenic model \cite{scot}.
In these model, scalar potential is given as 
\begin{eqnarray}
V_1&=&\lambda_1(\phi^\dagger\phi)^2+\kappa_S(S^\dagger S)^2 
+\kappa_{S\phi}(S^\dagger S)(\phi^\dagger\phi) +m_\phi^2\phi^\dagger\phi \nonumber\\
&+&\alpha(S^4 +S^{\dagger 4}) + \beta(S^2 +S^{\dagger 2} )\phi^\dagger\phi +m^2_S(S^\dagger S), 
\nonumber\\
V_2&=&\lambda_2(\eta^\dagger\eta)^2+\lambda_3(\phi^\dagger\phi)(\eta^\dagger\eta)
+\lambda_4(\phi^\dagger\eta)(\eta^\dagger\phi) 
+\frac{\lambda_5}{2}\left[(\eta^\dagger\phi)^2+{\rm h.c.}\right] \nonumber\\
&+&\kappa_{S\eta}(S^\dagger S)(\eta^\dagger\eta) + m_\eta^2\eta^\dagger\eta, 
\label{spot}
\end{eqnarray}
where $\phi$ is an ordinary SM doublet Higgs scalar and $\eta$ is an additional doublet scalar called 
an inert doublet scalar. $V_1$ is the one for the type-I seesaw model, 
and $V_1+V_2$ is the one 
for the scotogenic model. A second line of $V_1$ causes the spontaneous $CP$ 
violation addressed above.\footnote{We note that this $CP$ violation does not cause a problematic 
contribution to the EDM of quarks and charged leptons.}

In the type-I seesaw model where $\varphi$ corresponds to the ordinary 
doublet Higgs scalar $\phi$,  $\Lambda_k$ is represented as 
\begin{equation}
\Lambda_k=\frac{\langle\phi\rangle^2}{M_{N_k}}.
\label{type1}
\end{equation} 
If $M_{N_k}$ takes a value of $O(10^{13})$ GeV for example, $\Lambda_k$ is found to be $O(1)$ eV.
On the oher hand, in the scotogenic model where the neutrino mass is generated 
radiatively through a one-loop diagram, the role of $\varphi$ in eq.~(\ref{yukawa}) is played 
by the inert doublet scalar $\eta$ with an odd charge under imposed $Z_2$ symmetry.
Since all the SM contents and $S$ are assumed to have its even charge,
$N_k$ has to have the odd charge.
If this symmetry is assumed to be unbroken and then $\eta$ cannot have a VEV, 
neutrino mass is forbidden at tree level.
However, $\eta$ can have interaction terms with the ordinary Higgs scalar $\phi$ as found 
from $V_2$ in eq.~(\ref{spot}). 
Since the VEV of $\phi$ is mediated to neutrino sector through these couplings, 
$\Lambda_k$ is caused through one-loop diagrams as
\begin{equation}
\Lambda_k=\frac{\lambda_5\langle\phi\rangle^2}{8 \pi^2M_{N_k}}
  \left[\frac{M_{N_k}^2}{M_\eta^2-M_{N_k}^2}
    \left(1+\frac{M_{N_k}^2}{M_\eta^2-M_{N_k}^2}
    \ln\frac{M_{N_k}^2}{M_\eta^2}\right) \right],
\label{scot}
\end{equation}
where $M_\eta$ is an averaged mass of neutral components of $\eta$.
If $M_{N_k} \gg M_\eta$ is satisfied, $\Lambda_k$ can be approximated as
\begin{equation}
\Lambda_k\simeq \frac{\lambda_5\langle\phi\rangle^2}{8 \pi^2M_{N_k}}\ln\frac{M_{N_k}^2}{M_\eta^2}.
\label{mapp}
\end{equation}
It shows that the right-handed neutrino mass $M_{N_k}$ could be much 
smaller than the one in the type-I seesaw model to cause a same value for 
the scale $\Lambda_k$ as long as $|\lambda_5|$ is much smaller than 1.
In addition to this point, it should be noted that the model can have an other interesting feature
compared with the ordinary type-I seesaw model.  
A lightest neutral component of $\eta$ can play a role of dark matter whose stability 
is guaranteed by the imposed $Z_2$ symmetry \cite{scotdm,ks}.\footnote{It is useful to note that
dark matter physics gives a lower bound for $|\lambda_5|$ \cite{inel,ks}. }

We focus our study on the model in which the lightest neutrino mass is zero.
 It is well-known that such a possibility can be realized in the above models with two right-handed 
neutrinos only. We assume such a case at first and extend it after that. 
We also adopt a tribimaximal mixing \cite{otribi} for a neutrino mixing matrix $U_\nu$,
which is a rather good phenomenological mixing matrix in the neutrino sector. 
Unfortunately, the tribimaximal mixing 
is known to cause an experimentally disfavored mixing angle $\theta_{13}=0$. 
However, its nonzero value in the PMNS matrix \cite{pmns} could be obtained with 
the help of nontrivial flavor mixing in a charged lepton sector \cite{charged}.

We present a possible phenomenological construction of the PMNS matrix 
in the present model here. 
The PMNS matrix is described as $V=U_e^\dagger U_\nu$ where $U_e$ 
is a diagonalization matrix of a charged lepton mass matrix ${\cal M}_e$. It is fixed by solving a relation  
$U_e^\dagger{\cal M}_e{\cal M}_e^\dagger U_e=({\cal M}_e^{\rm diag})^2$ where 
${\cal M}_e^{\rm diag}$ is a diagonalized charged lepton mass matrix and $U_\nu$ is taken as a 
tribimaximal mixing matrix.
The matrix $U_e$ is expected to be almost diagonal like a CKM matrix \cite{ckm} since hierarchical 
masses of the charged leptons suggest ${\cal M}_e$ to be nearly diagonal. 
Characteristic structure of the PMNS matrix $V$ is considered 
to be determined by the mixing matrix $U_\nu$ in the neutrino sector. 
If we assume $U_e$ to be written with the CKM matrix $V_{\rm CKM}(\equiv U_u^\dagger U_d)$,
we find that $V_{13}$ takes a nonzero value  $(V_{\rm CKM})_{21}/\sqrt 2 \simeq 0.15$, 
which is required by the neutrino oscillation data.
This fact encourages us to take a phenomenological assumption  
such that the PMNS matrix $V$ is given as $V=V_{\rm CKM}^T U_\nu$.

This assumption might be supported from the flavor structure 
of Yukawa couplings in $SU(5)$ grand unification. 
Its multiplet structure ${\bf 10}_L =(Q_L,~u_R^c,~e_R^c)$ and ${\bf 5}_L^\ast=( d_R^c, ~\ell_L)$ 
results in a mass matrix relation between charged leptons and down-type quarks such 
as ${\cal M}_e={\cal M}_d^T$ \cite{su5}. When they are hermitian, 
their diagonalization matrices are expected to satisfy $U_e=U_d^\ast$.
It suggests $U_e\simeq V_{\rm CKM}^\ast$ if we take a basis for which a mass matrix 
of the up-type quarks is diagonal. 
Applying this assumption to the present model, we have
\begin{equation}
V=\left(\begin{array}{ccc}0.89 & 0.43 & 0.15\\
0.23 & 0.72 & 0.66\\
0.39 & 0.55 & 0.74 \\ \end{array}\right), 
\label{ckm}
\end{equation}
where all elements are represented by their absolute values, and
the Jarlskog invariant takes $J=4.1\times 10^{-4}$.\footnote{This result is not changed 
by the Majorana phases since $J$ does not generally depend on them as found from its definition 
$J={\rm Im}(V_{12}V_{23}V^\ast_{13}V^\ast_{22})$ \cite{jar}.}
More than half  of the $V$ components is included in the 3$\sigma$ range of the 3$\nu$ global fit 
\cite{3nu}. A possible larger value of the Jarlskog invariant $J$ is discussed in Appendix.

In the following part, our study is focused on the case featured by a zero 
mass eigenvalue and a tribimaximal mixing matrix $U_\nu$.
We suppose that the neutrino mass matrix ${\cal M}_\nu$ is diagonalized by the 
tribimaximal mixing matrix $U_\nu$ as
\begin{eqnarray}
&&U_\nu^T{\cal M}_\nu U_\nu={\cal M}_\nu^{\rm diag}, \nonumber \\
&&{\cal M}_\nu^{\rm diag}=(0,m_2,m_3) \quad {\rm for~NO}, \qquad
{\cal M}_\nu^{\rm diag}=(m_1,m_2,0) \quad {\rm for~IO}.
\label{zerom}
\end{eqnarray}
The matrix $U_\nu$ can be expressed as
\begin{equation}
U_\nu=\left(\begin{array}{ccc}\frac{2}{\sqrt 6} & \frac{1}{\sqrt 3} & 0\\
\frac{-1}{\sqrt 6} & \frac{1}{\sqrt 3} & \frac{1}{\sqrt 2}\\
\frac{1}{\sqrt 6} & \frac{-1}{\sqrt 3} & \frac{1}{\sqrt 2}\\ \end{array}\right)
\left(\begin{array}{ccc}e^{i\alpha_1} & 0 & 0\\
0 & e^{i\alpha_2} & 0\\
0 & 0 &e^{i\alpha_3} \\ \end{array}\right).
\label{unu}
\end{equation}
In general, since one of $\alpha_k$, which may be chosen as
$\alpha_1$, can be removed from the PMNS matrix by field redefinition, only two phases 
$\alpha_{2,3}-\alpha_1$ become physical. However, in the case of $m_1=0$ for example, 
$\alpha_1$ can be fixed freely. As a result, no phase remains as a physical 
one in the case $\alpha_2=\alpha_3$ although one phase can be physical in the case 
$\alpha_2\not=\alpha_3$.

By solving eq.~(\ref{zerom}), we can derive conditions for $h^\nu_{\alpha k}$.
In fact,  in the NO case, if they are defined as \cite{tribia,tribib}
\begin{equation}
h_{e2}^\nu=h_{\mu 2}^\nu=-h_{\tau 2}=h_2; \quad 
h_{e3}^\nu=0,~  h_{\mu 3}^\nu=h_{\tau 3}^\nu=h_3,
\label{tribi}
\end{equation}
 ${\cal M}_\nu$ is found to be expressed as
\begin{equation}
{\cal M}_\nu=\frac{1}{3}m_2e^{2i\alpha_2}\left(\begin{array}{ccc}1&1&-1\\ 1&1&-1\\ -1&-1&1\\\end{array}
\right)+\frac{1}{2}m_3e^{2i\alpha_3}\left(\begin{array}{ccc}0&0&0\\ 0&1&1\\ 0&1&1\\\end{array}
\right),
\end{equation}
where $m_{2,3}$ and $\alpha_{2,3}$ are represented as
\begin{equation}
m_2=3h_2^2\Lambda_2, \qquad m_3=2h_3^2\Lambda_3\ \qquad \alpha_{2,3}=\frac{\gamma}{2}.
\end{equation}
On the other hand, in the IO case, if Yukawa couplings $h^\nu_{\alpha k}$ satisfy
\begin{equation}
\frac{1}{2}h_{e1}^\nu=- h_{\mu 1}^\nu=h_{\tau 1}^\nu=h_1; \quad 
h_{e2}^\nu=h_{\mu 2}^\nu=-h_{\tau 2}=h_2,
\label{tribii}
\end{equation}
 ${\cal M}_\nu$ can be expressed as
\begin{equation}
{\cal M}_\nu=\frac{1}{6}m_1e^{2i\alpha_1}\left(\begin{array}{ccc}4&-2&2\\ -2&1&-1\\ 2&-1&1\\\end{array}
\right)+\frac{1}{3}m_2e^{2i\alpha_2}\left(\begin{array}{ccc}1&1&-1\\ 1&1&-1\\ -1&-1&1\\\end{array}
\right),
\end{equation}
where $m_{1,2}$ and $\alpha_{1,2}$ are represented as
\begin{equation}
 m_1=6h_1^2\Lambda_1, \qquad m_2=3h_2^2\Lambda_2, \qquad \alpha_{1,2}=\frac{\gamma}{2}.
\end{equation}
All the Majorana phases in eq.~(\ref{unu}) are removed from the PMNS 
matrix by the field redefinition and becomes unphysical because of the reason addressed 
below eq.~(\ref{unu}).

Now we consider to extend it to a model with three right-handed neutrinos keeping 
a zero mass eigenvalue by introducing the third right-handed neutrino which has mass terms
$(M_j + \Delta M)\bar N_jN_j^c$ where $j=1$ for the NO and $j=3$ for the IO.
The absolute value of its mass $M_{N_j}$ and the phase $\tilde\gamma$ are expressed as
\begin{equation}
M_{N_j}=\sqrt{M_j^2+M_0^2+2M_jM_0\cos(\gamma-\xi)},
\quad
\tan\tilde\gamma=\frac{M_j\sin\gamma+M_0\sin\xi}{M_j\cos\gamma+M_0\cos\xi},
\label{tgam}
\end{equation}  
where $M_j=y_ju$ and  $\Delta M=M_0e^{i\xi}~~(\xi\not=\gamma)$.
It is crucial to note that this extension is necessary to make 
leptogenesis possible in this scenario as shown in the next section.
The extension can be realized in the NO by assuming flavor structure of the Yukawa 
coupling $h^\nu_{\alpha 1}$ 
to be the same as the one of $h^\nu_{\alpha 2}$ or $h^\nu_{\alpha 3}$. 
Thus, two possibilities can be considered in each ordering.
 
If we take the NO as an example here, required conditions and the change of 
eigenvalues can be expressed as
\begin{eqnarray}
&&{\rm (a)} ~h_{e1}^\nu=h_{\mu 1}^\nu=-h_{\tau 1}^\nu=h_1; \quad 
m_2=3h_1^2\Lambda_1e^{i(\tilde\gamma-\gamma)}+3h_2^2\Lambda_2,
\quad m_3=2h_3^2\Lambda_3, \nonumber \\
&&{\rm (b)}~h_{e1}^\nu=0,~  h_{\mu 1}^\nu=h_{\tau 1}^\nu=h_1; \quad m_2=3h_2^2\Lambda_2, \quad
m_3=2h_1^2\Lambda_1e^{i(\tilde\gamma-\gamma)}+2h_3^2\Lambda_3.
\label{h1cond}
\end{eqnarray} 
where $\alpha_{1,2,3}=\frac{\gamma}{2}$ is used.
One Majorana phase is found to remain as the physical phase.
Although $m_2$ or $m_3$ is modified in each case, the lightest mass eigenvalue $m_1$ 
is kept to be zero.  Thus, the difference of squared masses fixed by neutrino oscillation data 
can directly determine the absolute values of neutrino masses in the present model. 

The above formulas of mass eigenvalues show that they could be almost 
independent of $h_1$ as long as $h_1$ contribution is much smaller compared 
with the one from $h_{2,3}$.\footnote{Mass eigenvalues $m_2$ or $m_3$ in each case 
can have the similar order contribution from $h_1$ and $h_k~(k=2,3)$ when 
$h_1\simeq(y_1/y_k)^{1/2}h_k$ is satisfied. If $h_1$ is smaller than it, 
the supposed situation happens.
In the case (b), for example, if $h_1$ contribution to $m_3$ is smaller than the 
$1\sigma$ deviation of the experimental value,
$h_1$ should be $h_1<5.2\times 10^{-4}$ in the type-I seesaw model with $M_{N_1}=10^9$ GeV,
and $h_1<1.4\times 10^{-3}$ in the scotogenic model with $M_{N_1}=10^4$ GeV, 
$M_\eta=10^3$ GeV  and $|\lambda_5|=10^{-5}$.} 
In that case, Majorana phases could be almost negligible in the PMNS matrix.
Although $h_1$ is kept as a free parameter as a result of the existence of
a zero mass eigenvalue, $h_{2,3}$ can be definitely determined through the neutrino 
oscillation data. Under the assumption $h_1\ll h_{2,3}$, we find 
\begin{eqnarray}
&{\rm (a)}& m_2=3h_2^2\Lambda_2=0.0087~{\rm eV}, \qquad m_3=2h_3^2\Lambda_3=0.0503~{\rm eV},
\nonumber \\
&{\rm (b)}& m_2=3h_2^2\Lambda_2=0.0087~{\rm eV}, \qquad m_3=2h_3^2\Lambda_3=0.0503~{\rm eV}
\label{meigen}
\end{eqnarray}
by using the experimental values for the NO given in \cite{pdg}.  
Values of $h_2$ and $h_3$ can be fixed as
\begin{equation}
h_2= 9.8\times 10^{-4}\left(\frac{M_{N_2}}{10^{10}~{\rm GeV}}\right)^{1/2}, \qquad
h_3= 9.1\times 10^{-3}\left(\frac{M_{N_3}}{10^{11}~{\rm GeV}}\right)^{1/2},
\label{type1v}
\end{equation}
by using eq.~(\ref{type1}) for the type-I seesaw model, and 
\begin{eqnarray}
&&h_2= 2.8\times 10^{-3}\left(\frac{10^{-5}}{|\lambda_5|}\frac{M_{N_2}}{10^5~{\rm GeV}}\right)^{1/2}
\left(1+0.22\ln\frac{M_{N_2}}{10^5~{\rm GeV}}\right)^{-1/2}, \nonumber \\
&&h_3=2.2\times 10^{-2} \left(\frac{10^{-5}}{|\lambda_5|}\frac{M_{N_3}}{10^6~{\rm GeV}}\right)^{1/2}
\left(1+0.14\ln\frac{M_{N_3}}{10^6~{\rm GeV}}\right)^{-1/2}
\label{scotv}
\end{eqnarray}
by using eq.~(\ref{mapp}) for the scotogenic model with $M_\eta=1$~TeV.
In the IO, the extension to the three right-handed neutrinos and the estimation of 
neutrino Yukawa couplings can be done in the same way as the one in the NO.

The rather light right-handed neutrinos and the additional 
inert doublet scalar in the scotogenic model are expected to cause a new non-negligible 
contribution to lepton flavor violating process like $\mu\rightarrow e\gamma$ \cite{tribib,kms}.
Differently from the type-I seesaw model, the contribution caused by the right-handed neutrinos 
at one-loop level could be rather large since the new doublet scalar is introduced.
If we use the neutrino Yukawa couplings fixed in eqs.~(\ref{tribi}) and (\ref{scotv}), 
we can estimate it definitely. Branching ratio of the lepton flavor violating process 
$\ell_a\rightarrow\ell_b\gamma$ in the scotogenic model is expressed as \cite{lfv} 
\begin{equation}
B(\ell_a\rightarrow\ell_b\gamma)=\frac{3\alpha}{64\pi(G_F M_\eta^2)^2}
\left|\sum_{j=1}^3h_{\ell_aj}h_{\ell_bj}F_2\left(\frac{M_{N_j}}{M_\eta}\right)\right|^2,
\end{equation}
where $F_2(x)$ is defined as
\begin{equation}
F_2(x)=\frac{1-6x^2+3x^4+2x^6-6x^4\ln x^2}{6(1-x^2)^4}.
\end{equation}
We easily find that this branching ratio is much smaller than a value expected to be reached
by future experiments as long as we suppose that the DM is the lightest neutral 
component of $\eta$ and the baryon number asymmetry is generated through 
the thermal leptogenesis due to the decay of $N_1$ \cite{ks}.

Imaginary part of the same one-loop diagram with electrons in both external lines causes 
the electric dipole moment of  electron through a Majorana phase of the right-handed neutrino
induced by the $CP$ violation of the singlet scalar \cite{vec}. It can be estimated as
\begin{equation}
d_e/e =\sum_{j=1}^3\frac{h_{ej}^2m_e}{16\pi^2M_\eta^2}
F_2\left(\frac{M_{N_j}}{M_\eta}\right)\sin(\gamma-\tilde\gamma). 
\end{equation} 
It predicts $d_e/e=O(10^{-36})$ cm for the Yukawa couplings given in eq.(\ref{scotv}) even if
the maximal $CP$ violation is assumed. This value is much smaller than the present experimental 
bound \cite{pdg}. 

The above result fixes the absolute values of neutrino mass definitely.
By using eq.~(\ref{meigen}) and the best fit values of the elements of the PMNS matrix given 
in \cite{pdg}, we can calculate two effective masses relevant to two observables 
in addition to the sum of neutrino masses, which is given as $\sum_j m_j=0.059$~eV in the NO, 
and $\sum_j m_j=0.10$~eV in the IO. 
One of them is the effective mass for the neutrinoless double $\beta$ decay, which is 
predicted to be  
\begin{equation}
m_{\beta\beta}=\left|\sum_{j=1}^3 V_{1j}^2m_j\right|= 0.0035~{\rm eV} ~~{\rm for~NO},
\quad  m_{\beta\beta}=\left|\sum_{j=1}^3 V_{1j}^2m_j\right|= 0.049~{\rm eV} ~~{\rm for~IO}.
\label{db}
\end{equation}
Another one is the effective neutrino mass determined by the $\beta$ decay and 
it is predicted as 
\begin{equation}
m_{\beta}=\sqrt{\sum_{j=1}^3 |V_{1j}|^2m_j^2}=0.0086~{\rm eV} ~~{\rm for~NO}, \quad
m_{\beta}=\sqrt{\sum_{j=1}^3 |V_{1j}|^2m_j^2}=0.049~{\rm eV} ~~{\rm for~IO}.
\end{equation}
Majorana phase has no effect on these results since small $h_1$ makes 
the relevant Majorana phase negligibly small as found from eq.~(\ref{h1cond}).
In the NO, the above results are much smaller than their present bounds \cite{katrin,kamzen}. 
Moreover, they seem to be too small to be reached in near future experiments for them.
On the other hand,  $m_{\beta\beta}$ given for the IO in eq.~(\ref{db}) suggests that 
the present bound for the neutrinoless double $\beta$ decay 
\cite{kamzen} may have excluded the model already.
In the following part, we confine our study to the NO.

\section{Constraints from baryon number asymmetry}
In the previous section, neutrino Yukawa couplings $h_{2,3}$ are shown to be fixed definitely 
in the present models.
Although $h_1$ is free from the neutrino oscillation data , it can be constrained by other 
phenomenological requirement.
This type of models make leptogenesis \cite{fy} work well and generate the baryon number 
asymmetry through the out-of-equilibrium decay of the lightest right-handed neutrino $N_1$.
Flavor structure defined by eqs.~(\ref{tribi}) and (\ref{h1cond}) which causes the tribimaximal mixing and a
zero mass eigenvalue could bring about crucial effects on leptogenesis. Taking account of it,
we derive constraints on $h_1$ from a few conditions relevant to thermal leptogenesis
and discuss feasibility of thermal leptogenesis. 

\subsection{Production of lepton number asymmetry}
Since lepton number is violated in the second term in eq.~(\ref{yukawa}), the $N_1$ decay
could cause the lepton number asymmetry as long as $CP$ symmetry is violated in eq.~(\ref{yukawa}).
$CP$ asymmetry $\varepsilon$ in the decay $N_1\rightarrow \ell_j\varphi^\dagger$ 
can be estimated by using the assumptions (\ref{tribi}) and (\ref{h1cond}) as 
\cite{cpasym}\footnote{It is useful to note that $\varepsilon$ is invariant under unitary 
transformation of $h^\nu_{\alpha k}$ as found from its expression and then it is
independent of the PMNS matrix.}
\begin{eqnarray}
\varepsilon&\equiv&\frac{\sum_\alpha[\Gamma(N_1\rightarrow \ell_\alpha\varphi^\dagger)-
\Gamma(N_1^c\rightarrow \bar\ell_\alpha\varphi)]}
{\sum_\alpha\Gamma(N_1\rightarrow \ell_\alpha\varphi^\dagger)} \nonumber \\
&=&\frac{1}{8\pi}\sum_{k=2,3}
{\rm Im}\left[ \frac{(\sum_\alpha h_{\alpha1}^{\nu \ast} h_{\alpha k}^\nu)^2}
{\sum_\alpha h_{\alpha 1}^{\nu\ast} h_{\alpha 1}^\nu}\right]
F\left(\frac{M_{N_k}^2}{M_{N_1}^2}\right)  
=\left\{\begin{array}{ll} 
{\rm (a)}&\frac{3}{8\pi}h_2^2F\left(\frac{M_{N_2}^2}{M_{N_1}^2}\right) \sin(\gamma-\tilde\gamma) \\
{\rm (b)}&\frac{2}{8\pi}h_3^2F\left(\frac{M_{N_3}^2}{M_{N_1}^2}\right)\sin(\gamma-\tilde\gamma), \\
\end{array}\right.
\label{asymp}
\end{eqnarray}
where $F(x)=\sqrt{x}[1-(1+x)\ln\frac{1+x}{x}]+\frac{\sqrt x}{1-x}$. 
Although Yukawa couplings $h_{\alpha k}^\nu$ are assumed to be real, 
the masses of the right-handed neutrinos have the complex phases $\tilde\gamma$ and $\gamma$. 
After redefining the right-handed neutrinos to make their masses real,
a factor $\sin(\gamma-\tilde\gamma)$ is induced in $\varepsilon$.
In the following study, we assume $\sin(\gamma-\tilde\gamma)=O(1)$ which can be easily 
realized in eq.~(\ref{tgam}). As such an example,  one may suppose
$M_1=M_0$, $\gamma=\frac{\pi}{4}$ and $\xi=-\frac{\pi}{2}$ for which eq.~(\ref{tgam})
gives $\sin(\gamma-\tilde\gamma)=0.92$, and we find $y_1\sim \frac{M_{N_1}}{\sqrt 2 u}$ in that case. 
This is adopted as a benchmark case in the following study.
Here, it is important to note that two right-handed neutrino model gives $\varepsilon=0$ for 
the $N_2$ decay irrelevantly to the $CP$ violation since eq.~(\ref{tribi}) results 
in $(\sum_\alpha h_{\alpha 2}h_{\alpha 3})^2=0$.
It gives us a motivation to extend the model to the one with three right-handed neutrinos 
although the former can explain the neutrino oscillation data.\footnote{We should 
note that leptogenesis is possible even in the model with only two right-handed neutrinos  
if the tribimaximal assumption is not adopted.}

As found from eq.~(\ref{asymp}), since the $CP$ asymmetry $\varepsilon$ is independent of 
the coupling $h_1$ under the assumed flavor structure, $h_1$ does not affect 
a value of $\varepsilon$. Even if $h_1$ takes a very small value, the magnitude of 
$\varepsilon$ can keep a value to generate sufficient lepton number asymmetry  
for the explanation of the baryon number asymmetry in the Universe.
The $CP$ asymmetry  $\varepsilon$ in the case (b) tends to take a larger value than one in the case (a) 
for neutrino Yukawa couplings given in eqs.~(\ref{type1v}) and (\ref{scotv}).
In the following part, we focus our study on the case (b).
If we apply eqs.~(\ref{type1v}) and (\ref{scotv}) to the formula (\ref{asymp}) for the case 
$M_{N_3}=10^2M_{N_1}$, we have
\begin{equation}
\varepsilon=-9.1\times 10^{-8}\left(\frac{M_{N_3}}{10^{11}~{\rm GeV}}\right), \qquad
\varepsilon\simeq -5.3\times 10^{-7}\left(\frac{10^{-5}}{|\lambda_5|}\frac{M_{N_3}}{10^6~{\rm GeV}}\right)
\end{equation}
for the type-I seesaw model and the scotogenic model, respectively. 
The $CP$ asymmetry $\varepsilon$  is found to take a value of $O(10^{-7})$ in both models 
if $M_{N_1}$ is fixed appropriately.

The lightest right-handed neutrino $N_1$ is expected to decay around temperature $T_D$
which is fixed by a condition $\Gamma_{N_1}^D=H(T_D)$. $\Gamma_{N_k}^D$ is a decay width 
of $N_k$ and $H(T)$ is the Hubble parameter. They are given as
\begin{equation}
\Gamma_{N_k}^D=\frac{c_kh_k^2}{8\pi}M_{N_k}\left(1-\frac{M_\varphi^2}{M_{N_k}^2}\right)^2, 
\qquad H(T)=\left(\frac{\pi^2}{90}g_\ast\right)^{1/2}\frac{T^2}{M_{\rm pl}},
\label{hub}
\end{equation}
where $c_k$ is a constant determined by the flavor structure of the model 
and fixed to $c_1=c_3=2$ and $c_2=3$. $g_\ast$ is relativistic degrees of freedom at 
temperature $T$ and $M_{\rm pl}$ is the reduced Planck mass. 
 The temperature $ T_D$ can be estimated by using this condition for the case 
$M_{N_1}\gg M_\varphi$ as
\begin{equation}
\frac{T_D}{M_{N_1}}=
2.4\times \left(\frac{h_1}{10^{-8+n/2}}\right)\left(\frac{10^n~{\rm GeV}}{M_{N_1}}\right)^{1/2}.
\label{decay}
\end{equation}
A small value of $h_1$ causes the late time decay of $N_1$ so that the washout processes of 
the lepton number asymmetry generated through the decay of $N_1$ 
could be ineffective when its substantial decay starts. 
If washout of the lepton number asymmetry caused by inverse decay of
the right-handed neutrinos and 2-2 scatterings mediated by $N_{2,3}$ freezes out at temperature 
$T_F$ and $T_D<T_F$ is satisfied, the generated asymmetry through the $N_1$ decay 
is not reduced.

Here, we discuss constraints on $h_1$ from the baryon number asymmetry.
For this purpose, we define number density of a particle species $i$ in the comoving volume as 
$Y_i\equiv \frac{n_i}{s}$ by using its number density $n_i$ and the entropy density $s$.
Lepton number asymmetry is expressed as $Y_L\equiv \frac{n_\ell-n_{\bar\ell}}{s}$,
where $n_\ell$ and $n_{\bar \ell}$ are the number density of leptons and antileptons, respectively.
If the washout processes for the generated lepton number asymmetry freeze out at the 
temperature $T_D$ where
the substantial $N_1$ decay occurs, the lepton number asymmetry could be roughly 
estimated as its maximum value which can be expressed as 
$Y_L=\varepsilon Y_{N_1}^{\rm eq }$ by using $\varepsilon$ in eq.~(\ref{asymp}). 
$Y^{\rm eq}_{N_1}$ is the equilibrium number density of $N_1$ as a relativistic particle and 
then $Y^{\rm eq}_{N_1}=O(10^{-3})$.
The required baryon number asymmetry can be generated for $Y_L=O(10^{-10})$ 
through a sphaleron process and then $|\varepsilon|$ should take a value of $O(10^{-7})$. 
Eq.~(\ref{asymp}) shows that $h_3>3\times 10^{-3}$ is required to realize it for $\frac{M_{N_3}}{M_{N_1}}>10$
since $\left|F\left(\frac{M_{N_3}^2}{M_{N_1}^2}\right)\right|<1.5\times 10^{-1}$ is expected there. 
From eqs.~(\ref{type1v}) and (\ref{scotv}), this condition is found to be satisfied in both models 
if values of $M_{N_3}$ and $\lambda_5$ are fixed appropriately.

\begin{figure}[t]
\begin{center}
\includegraphics[width=7cm]{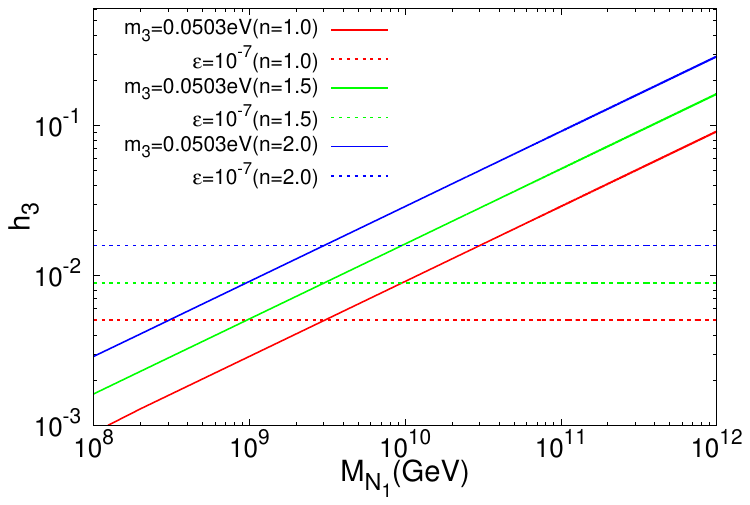}
\hspace*{5mm}
\includegraphics[width=7cm]{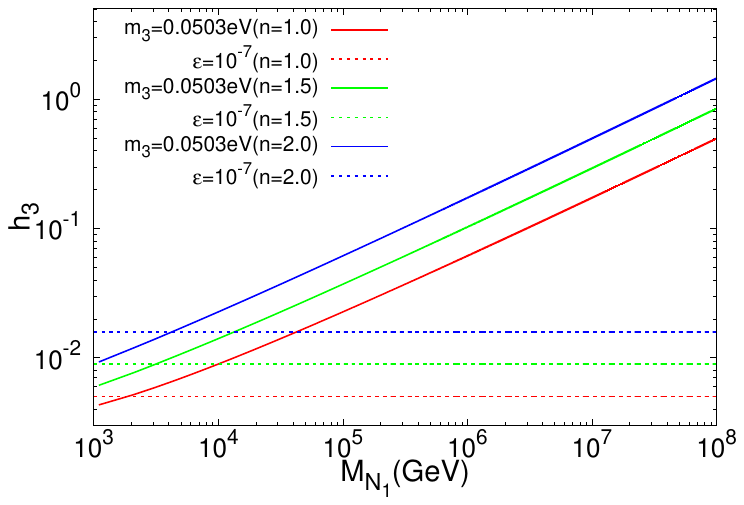}
\end{center}

\footnotesize{{\bf Fig.~1}~~Yukawa coupling $h_3$ determined through both a value of $m_3$ required 
by the neutrino oscillation data (solid lines) and the $CP$ asymmetry of the $N_1$ decay 
$\varepsilon=10^{-7}$ which is required for the successful leptogenesis (dotted lines). 
They are plotted as a function of $M_{N_1}(=10^{-n}M_{N_3})$ GeV.
The type-I seesaw model and the scotogenic model are shown in left and right panels, respectively.}
\end{figure}

To examine it in more definite way, we numerically solve a condition $\varepsilon=10^{-7}$ 
by using  eq.~(\ref{asymp}).\footnote{We should use this $\varepsilon$ value 
as a benchmark. It should be noted that a certain range around it can make leptogenesis 
work well.}
In Fig.~1, $h_3$ is plotted as a function of $M_{N_1}$ using dotted lines 
for three values of $M_{N_3}$ which are fixed through a relation  $M_{N_3}=10^nM_{N_1}~(n=1, 1.5, 2)$.
In the same panel, we also plot $h_3$ given by eqs.~(\ref{type1v}) and (\ref{scotv}) using solid lines. 
They are required for the explanation of the neutrino oscillation data.
Under the neutrino oscillation constraints, the required baryon number asymmetry can be generated 
in a region of $M_{N_1}$ where the solid line comes over the dotted line of the same color for each $n$.
In the type-I seesaw model shown in the left panel, we can see the leptogenesis works well 
for $M_{N_1}>3\times 10^9$ GeV independently of the $M_{N_3}$ value. 
This result is consistent with the well-known bound given in \cite{di}.
On the other hand, in the scotogenic mode shown in the right panel, we find successful 
leptogenesis could occur even for $N_1$ with TeV scale mass. 
An allowed minimum value of $M_{N_1}$ is found to have the weak dependence on $M_{N_3}$.
It is a different feature from the type-I seesaw model, which is caused by a loop effect. 

Finally, we estimate a range of $h_1$ required from necessary conditions for leptogenesis 
quantitatively.
Since the $N_1$ decay has to occur in the out-of-equilibrium situation and then 
$T_D~{^<_\sim}~ M_{N_1}$ should be satisfied, we find by using  eq.~(\ref{decay}) 
that $h_1$ should take a value in a range\footnote{This bound is consistent with the assumption 
for an $h_1$ value discussed in the footnote d.}  
\begin{equation}
h_1~{^<_\sim}~4.2\times 10^{-9+{n/2}}\left(\frac{M_{N_1}}{10^n~{\rm GeV}}\right)^{1/2}.
\label{h1cond1}
\end{equation}
On the other hand, the generated lepton number asymmetry has to be converted to 
the baryon number asymmetry through the sphaleron process, which is considered to 
be in the thermal equilibrium at temperature higher than 100 GeV \cite{spha}.  
This requires the $N_1$ decay to occur at higher temperature than it.
This imposes $T_D~{^>_\sim}~100$ GeV, which can be translated to a condition for $h_1$ as
\begin{equation}
h_1~{^>_\sim}~4.2\times 10^{-7-{n/2}}\left(\frac{10^n~{\rm GeV}}{M_{N_1}}\right)^{1/2}.
\label{h1cond2}
\end{equation}

In the type-I seesaw model, $M_{N_1}>3\times 10^9$ GeV should be satisfied from 
Fig.~1 and then eqs.~(\ref{h1cond1}) and (\ref{h1cond2}) suggest that $h_1$ can
take a value in a wide range 
$2.3\times 10^{-11}~{^<_\sim}~h_1~{^<_\sim}~2.3\times 10^{-4}$ 
for this minimum value of $M_{N_1}$. 
We can expect that leptogenesis occurs successfully for an $h_1$ value around the upper 
bound of this range. For such a value of $h_1$, $N_1$ is expected to be generated 
in the thermal bath through the inverse decay. 
On the other hand, in the scotogenic model with $M_\eta=O(1)$~TeV,
if we adopt a lower value  $M_{N_1}=10^4$ GeV allowed in Fig~1, we find that $h_1$ should be 
in a range $4.2\times10^{-9}~{^<_\sim}~h_1~{^<_\sim}~4.2\times 10^{-7}$.\footnote{If one 
thinks that $h_1$ is unnaturally small, it may be useful to note that 
$h_1/h_3$ is comparable to the ratio of the Yukawa couplings of up-quark and top-quark. }  
The range is rather narrow compared to the one in the type-I seesaw model with a heavy $N_1$.
Moreover, $h_1$ seems to be too small to generate $N_1$ in the thermal bath. 
In the next parts, we discuss both thermal production of $N_1$ and washout of the generated 
lepton number asymmetry assuming that $h_1$ is contained in the above range.

\subsection{Thermal production of $N_1$}
Since $h_1$ is assumed to be small in the present study, 
$N_1$ may not be generated sufficiently to reach the thermal equilibrium by the neutrino 
Yukawa coupling $h_1$. In that case, some other certain interaction in the model 
has to play a role for its generation instead of the coupling $h_1$. 
A scattering process mediated by $S$ could contribute to it in the present models.
 
In the thermal leptogenesis in both the type-I seesaw model and the scotogenic model, 
thermal right-handed neutrinos are generally considered to be produced by 
neutrino Yukawa couplings $h^\nu_{\alpha k}$. 
Since $\ell_\alpha$ and $\varphi$ are considered to be in the thermal bath through 
the SM interactions, they generate $N_k$ thermally through the inverse decay 
$\ell_\alpha\varphi^\dagger\rightarrow N_k$ 
and scattering processes $\ell_\alpha\bar\ell_\beta\rightarrow N^c_jN_k$ and 
$\varphi\varphi^\dagger\rightarrow N^c_jN_k$.  Since the latter are higher order than
the inverse decay, the inverse decay is considered to be dominant among them 
at the temperature $T>M_{N_k}$ where it is not suppressed energetically.
We can estimate the magnitude of neutrino Yukawa couplings required to yield $N_k$ 
 in the thermal equilibrium through the inverse decay using a condition 
$\Gamma_{N_k}^D \ge H(M_{N_1}/2)$. 
$\Gamma_{N_k}^D$ and $H(T)$ are presented in eq.~(\ref{hub}). 
In the case $M_\varphi\ll M_{N_1}$, we find that this condition is satisfied for 
\begin{equation}
h_k~{^>_\sim}~3.0\times10^{-9+n/2}c_k^{-1/2}\left(\frac{M_{N_k}}{10^n~{\rm GeV}}\right)^{1/2}.
\label{ubh1}
\end{equation} 
Neutrino Yukawa couplings which explain the neutrino oscillation data should also satisfy this 
condition to guarantee the existence of $N_k$ in the thermal bath. 
In the present models, it can be confirmed for $N_{2,3}$ whose Yukawa couplings $h_{2,3}$ 
are fixed by eqs.~(\ref{type1v}) and (\ref{scotv}). 
On the other hand, $N_1$ is not expected to be in the thermal equilibrium unless $h_1$ takes a value
in the neighborhood of the upper bound in the region given in eq.~(\ref{h1cond1}) at least.

In the present models, right-handed neutrino mass is assumed to be induced from 
the VEV of the singlet scalar $S$ through the second term in eq.~(\ref{yukawa}). 
We should note that the same interaction can cause scattering $N_kN_k \rightarrow N_1N_1$.
Since $N_{2,3}$ can be in the thermal equilibrium through the inverse decay 
as discussed above, this process could contribute to the production of $N_1$ 
in the thermal bath even if $h_1$ is small not to fulfill the condition (\ref{ubh1}). 
Cross section of the scattering $N_kN_k \rightarrow N_1N_1$ mediated by the singlet scalar 
$S$ is estimated as
\begin{equation}
\sigma v=\frac{y_1^2y_k^2}{32 \pi s}\sqrt{1-\frac{4M_{N_1}^2}{s}}
\frac{(s-2M_{N_1}^2)(s-2M_{N_k}^2)}{(s-m_S^2)^2},
\end{equation}
where $v$ is relative velocity of $N_k$'s and $s$ is energy of the center of mass system.
$m_S$ is the mass of $S$ which is assumed to satisfy $m_S \ll M_{N_k}$, and then $S$ cannot 
decay to a pair of $N_k$.\footnote{If $S$ can decay to a pair of $N_k$, leptogenesis could 
occur nonthermally \cite{nonth} which is not considered here.
This assumption requires that a quartic coupling of $S$ 
satisfies $\sqrt\kappa_S\ll y_k$. If a component $S$ plays a role of inflaton \cite{sinf}, 
CMB data requires $\kappa_S=O(10^{-8})$ which could justify this inequality.}

\begin{figure}[t]
\begin{center}
\includegraphics[width=7cm]{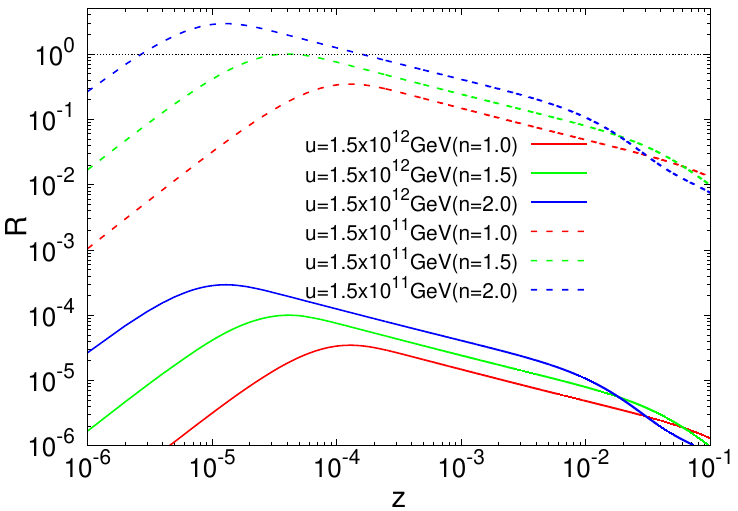}
\hspace*{5mm}
\includegraphics[width=7cm]{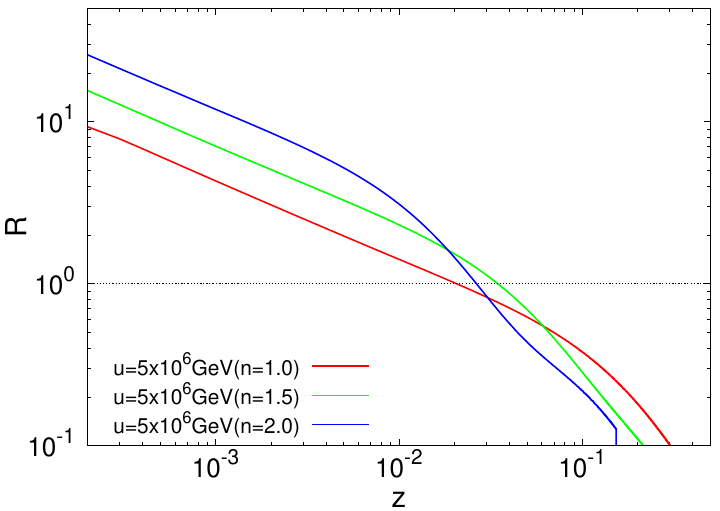}
\end{center}
\vspace{-2mm}

\footnotesize{{\bf Fig.~2}~~Ratio $R$ of the reaction rate $\Gamma(z)$ of the $N_1$ production 
due to scattering mediated by $S$ to the Hubble parameter $H(z)$. It is plotted 
as a function of $z(\equiv M_{N_1}/T)$. 
$M_{N_3}=10^nM_{N_1}$ and $M_{N_2}=10^{n/2}M_{N_1}$ are assumed.
Left panel : the type-I seesaw model with $M_{N_1}=3\times 10^9$ GeV, 
$u=1.5\times 10^{12}$ GeV and $h_1=10^{-4}$ (solid lines), and $M_{N_1}=3\times 10^9$ GeV, 
$u=1.5\times 10^{11}$ GeV and $h_1=10^{-4}$ (dotted lines). 
Right panel : the scotogenic model with $M_{N_1}=10^4$ GeV, $u=5\times 10^6$ GeV
and $h_1=10^{-8}$. }
\end{figure}

We examine the effectiveness of the $N_1$ production through this process in each model.
Reaction rate $\Gamma$ of the process at temperature $T~ (>M_{N_k})$ can be estimated as 
$\Gamma(T)=\langle\sigma v\rangle n(T)$ by using the thermally averaged cross section 
$\langle\sigma v\rangle$ and the number density of a target $n(T)$.
If $\Gamma(T)\ge H(T)$ is satisfied for the Hubble parameter $H(T)$,
this process is expected to reach the equilibrium at temperature $T$. 
Since this condition can be roughly estimated as $M_k<T~{^<_\sim}~ 10^{16}y_1^2y_k^2$ GeV,
the $N_1$ production through this scattering tends to be suppressed by a Boltzmann 
factor in models with heavy right-handed neutrinos like the type-I seesaw model.
In fact, if we take account of $M_1>3\times 10^9$ GeV and assume $M_{N_k}=10^nM_{N_1}$,
this condition requires $y_k >1$ for $n=2$, for example.
It suggests that this scattering cannot be so effective unless $y_k$ takes a rather large value
in that kind of model. 

In Fig.~2, we plot $R=\Gamma(z)/H(z)$ for this scattering process as a function 
of $z(\equiv M_{N_1}/T)$ in the left panel for the type-I seesaw model and in the right panel 
for the scotogenic model. In this calculation, we assume $M_{N_3}=10^nM_{N_1}$ and 
$M_{N_2}=10^{n/2}M_{N_1}$, and values of Yukawa couplings $y_k$ are fixed by using the VEV 
$u$ as $y_{2,3}=M_{N_{2,3}}/u$ and $y_1=M_{N_1}/(\sqrt 2u)$.  
Concerning the solid line for each $n$, $u$ is fixed for the Yukawa coupling 
$y_k$ to take the same value in both models.  As a reference for a larger $y_k$ case 
in the type-I seesaw model, $R$ is also plotted for $u=1.5 \times 10^{11}$ GeV by the dotted lines.  
The figures shows that this scattering process is very effective for the $N_1$ production 
in the scotogenic model in which $N_1$ can be rather light.
On the other hand, it cannot play a substantial role in the type-I seesaw model unless the Yukawa 
coupling $y_k$ takes a larger value of $O(1)$.
However, as long as the coupling $h_1$ is fixed to an appropriate value around the upper bound 
in the range given in eq.~(\ref{h1cond1}), $N_1$ can be produced only by the 
coupling $h_1$ and leptogenesis works well. 
In the scotogenic model, this situation is similar in a heavy $N_1$ case like $M_{N_1}>10^9$~GeV \cite{ks}. 
Crucial change could be caused in a small $M_{N_1}$ case such as $M_{N_1}\ll 10^9$~GeV.
By introducing the interaction which fixes the origin of the right-handed neutrinos mass,  
the scattering mediated by $S$ could produce $N_1$ in the thermal bath
even in the case where the coupling $h_1$ cannot produce the sufficient $N_1$ thermally.
It might be recognized as an another dedicated feature of the model, which can present 
the origin of dark matter.

\subsection{Washout of lepton number asymmetry}
The generated lepton number asymmetry can be washed out by the inverse decay
$\ell_\alpha\varphi\rightarrow N_k$, and the 2-2 scatterings 
$\ell_\alpha\varphi^\dagger\rightarrow\bar\ell_\beta\varphi$ and 
$\ell_\alpha\ell_\beta\rightarrow \varphi\varphi$ which are mediated by the right-handed 
neutrino $N_k$ through the neutrino Yukawa couplings.
Since the washout effect becomes crucial at $T~{^<_\sim}~M_{N_1}$ where the out-of-equilibrium 
decay of $N_1$ occurs and the inverse decay is suppressed by the Boltzmann factor,
it is considered to be dominantly caused by the 2-2 scatterings.
If we take account of the assumed flavor structure of neutrino Yukawa couplings in eqs.~(\ref{tribi})
and (\ref{h1cond}), their cross section is calculated as
\begin{eqnarray}
\sigma v&=&\sum_{k} \frac{(c_kh_k^2)^2M_{N_k}^2}{16\pi}\left[\frac{1}{s}\left\{(s^2-4sM_\varphi^2)^{1/2}
\frac{2}{(M_\varphi^2-M_{N_k}^2)^2 +sM_{N_k}^2}\right.\right. \nonumber \\
&+&\left. \frac{2}{2M_\eta^2-2M_{N_k}^2-s}\ln\left(\frac{2M_\eta^2-2M_{N_k}^2-s+(s^2-4sM_\eta^2)^{1/2}}
{2M_\eta^2-2M_{N_k}^2-s-(s^2-4sM_\eta^2)^{1/2}}\right)\right\}
\nonumber\\
&+&\frac{1}{(s+M_\varphi^2)}\left\{ 
\frac{(s-M_{N_k}^2)^2(s-M_\varphi^2)^2}
{s((s-M_{N_k}^2)^2 +M_{N_k}^2\Gamma_{N_k}^2)^2}
\right.\nonumber \\
&+&\frac{4(s-M_{N_k}^2)}{(s-M_{N_k}^2)^2 +M_{N_k}^2\Gamma_{N_k}^2} 
\left(1+\frac{(2M_\varphi^2-M_{N_k}^2-s)s}{(s-M_\varphi^2)^2}
\ln\left(\frac{(2M_\varphi^2-M_{N_k}^2-s)s}{M_\varphi^4-s M_{N_k}^2}\right)\right) \nonumber \\
&+&\left.\left.\frac{2s}{s M_{N_k}^2-M_\varphi^4} + 
\frac{2s}{(s-M_\varphi^2)^2}
\ln\left(\frac{M_\varphi^4-s M_{N_k}^2}{(2M_\varphi^2 -M_{N_k}^2-s)s}\right)\right\}\right],  
\end{eqnarray}
where $\Gamma_{N_k}^D$ is the decay width of the right-handed neutrino $N_k$ given in eq.~(\ref{hub}).
Reaction rate of the scattering can be given by $\Gamma(T)=\langle\sigma v\rangle n(T)$ 
using thermally averaged cross section $\langle\sigma v\rangle$ and the number density $n(T)$ 
of relativistic particles. 
Since freezeout of the process is expected to occur when $\Gamma(T)<H(T)$ is satisfied,
the washout effect is expected to be negligible at the temperature lower than $T_F$ defined by
$\Gamma(T_F)=H(T_F)$.

\begin{figure}[t]
\begin{center}
\includegraphics[width=7cm]{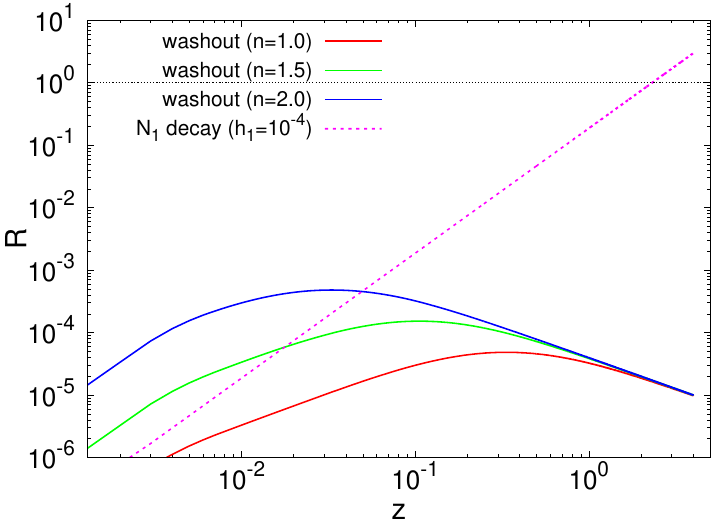}
\hspace*{5mm}
\includegraphics[width=7cm]{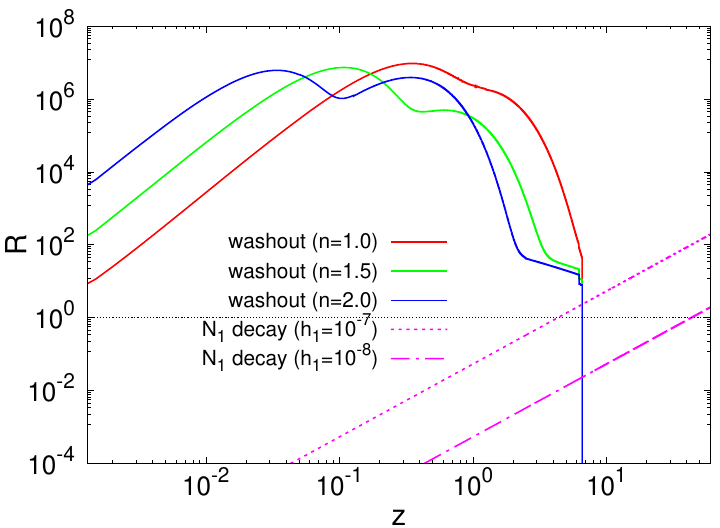}
\end{center}
\vspace*{-2mm}

\footnotesize{{\bf Fig.~3}~~Ratio $R$ between the reaction rate $\Gamma(z)$ of 
the washout processes of the lepton number asymmetry and 
the Hubble parameter $H(z)$. It is plotted as a function of $z(\equiv M_{N_1}/T)$ 
for the type-I seesaw model with $M_{N_1}=3\times 10^9$ GeV (left panel) 
and the scotogenic model with $M_{N_1}=10^4$ GeV (right panel).
Mass of $N_{2,3}$ is fixed to three values defined by $n$ shown in each panel.
Freezeout temperature $T_F$ of the washout process is read off from the condition $R=1$.
$R$ of the $N_1$ decay is also plotted as a reference.}
\end{figure}

In Fig.~3, using the neutrino Yukawa couplings given in eqs.~(\ref{type1v}) and (\ref{scotv}) 
we plot $R=\Gamma(z)/H(z)$ for the washout processes for three values of $M_{N_{2,3}}$
defined as $M_{N_3}=10^nM_{N_1}$ and $M_{N_2}=10^{n/2}M_{N_1}$ again. 
In the type-I seesaw model with $M_{N_1}=3\times 10^9$ GeV plotted in the left panel, 
the washout processes are found to be sufficiently suppressed not to affect the generation 
of the baryon number asymmetry at the interesting region $z~{^>_\sim}~1$. 
In the scotogenic model with $M_{N_1}=10^4$ GeV which is plotted in the right panel, 
the washout effect is still strong at $z\simeq 1$ where the out-of-equilibrium 
decay of $N_1$ is usually expected to be substantial.\footnote{This situation is generally expected 
also in the model with no zero mass eigenvalue where $h_1$ gives non-negligible contribution 
to the neutrino mass.} 
However, since the $N_1$ production due to the coupling $h_1$ needs not be taken into account, 
$h_1$ can be  small enough within the allowed range fixed by eqs.~(\ref{h1cond1}) 
and (\ref{h1cond2}) to make the $N_1$ decay delay sufficiently. 
In that case, the decay occurs at a much lower temperature than $T_F$ and then the 
washout of the generated lepton number asymmetry can be neglected. 
In fact, $T_D<T_F$ is found in the figure where $h_1=10^{-8}$ is adopted from the allowed range
for $M_{N_1}=10^4$ GeV. Thus, the lepton number asymmetry supposed in the previous 
section can be directly applicable without considering the influence of the washout effect.
These observations can be directly examined by solving Boltzmann equations 
for the relevant particles. In Fig.~4, we display the time evolution of the number density of the relevant particles in the comoving volume in the scotogenic model defined by $n=2.0$ in Fig.~2 and 3.
We can confirm the validity of the above observations through this figure.

\begin{figure}[t]
\begin{center}
\includegraphics[width=8cm]{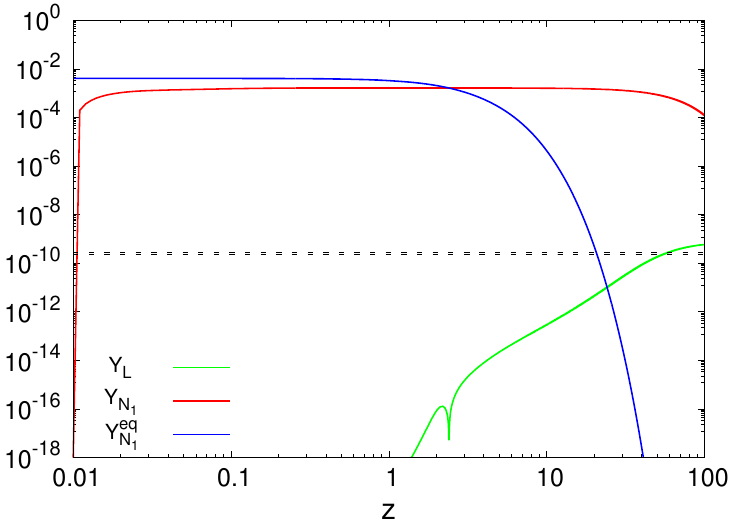}
\end{center}
\vspace*{-2mm}

\footnotesize{{\bf Fig.~4}~~Evolution of number density of particle $i$ in a comoving volume 
$Y_i$ in the scotogenic model. 
It is derived as a solution of Boltzmann equations and displayed as a function 
of $z(\equiv \frac{M_{N_1}}{T})$. $Y_L$ and $Y_{N_1}^{\rm eq}$ represent the lepton number asymmetry 
and an equilibrium value of $Y_{N_1}$, respectively. 
Horizontal dotted lines express values required to explain the baryon number in the Universe.} 
\end{figure}

In the scotogenic model, the assumption of the existence of a zero mass eigenvalue 
makes $h_1$ free from the neutrino oscillation constraints.
In addition to it, introduction of the interaction between $N_k$ and $S$ which gives the 
origin of the right-handed neutrino mass allows $h_1$ to take a very small value,
which makes the generated lepton number asymmetry free from the washout effects. 
These make low scale leptogenesis possible. Moreover, the late decay of $N_1$ caused 
by the small $h_1$ can affect the dark matter abundance. It makes $N_1$ a common progenitor 
of baryon number asymmetry in the Universe and dark matter as suggested in \cite{commonp}.
On the other hand, in the type-I seesaw model, the generation of sufficient lepton 
number asymmetry could be guaranteed for $M_{N_1}>3\times 10^9$~GeV 
since $N_1$ is sufficiently produced in the thermal bath through the neutrino Yukawa 
couplings which do not cause the disastrous washout of the lepton number asymmetry. 
It is consistent with the well-known lower bound of $M_{N_1}$ for the successful leptogenesis 
\cite{di}. The existence of a zero mass eigenvalue seems not to cause noticeable impact
on the leptogenesis. 

\section{Summary}
The absolute value of neutrino mass is not fixed still now. However, recent cosmological 
observations give a stringent constraint. From this viewpoint, the existence of a zero 
mass eigenvalue of neutrino seems to be a noticeable and realistic possibility. 
Model construction of such a feature is considered to be an interesting subject now. 
In this paper, we consider flavor structure of neutrino Yukawa 
couplings which can cause both a zero mass eigenvalue and a promising PMNS matrix.
We apply it to both the type-I seesaw model and the scotogenic model, and
estimate neutrino Yukawa couplings based on neutrino oscillation data.
The analysis on the mass ordering shows that the IO may have already excluded through 
the present bound of the neutrinoless double $\beta$ decay.
In the NO case, the predicted values for the effective mass of both the $\beta$ decay and 
the neutrinoless double $\beta$ decay are much smaller than their present bounds. 
Unfortunately, it seems to be too small to be reached in near future experiments.

We also study the neutrino Yukawa coupling of the lightest right-handed neutrino which 
cannot be fixed through the neutrino oscillation data and could take a very small value.
We suggest through this study that an interesting possibility appears in the leptogenesis
if we introduce the interaction between the right-handed neutrinos and a singlet scalar
which can give the origin of the right-handed neutrino mass through the VEV of 
the singlet scalar. In the scotogenic model, this interaction could play a crucial role to produce 
the lightest right-handed neutrinos in the thermal bath.
It makes low scale leptogenesis feasible since the model presents a circumstance in which
the washout of the generated lepton number asymmetry is negligible.
Although neutrino models with a zero mass eigenvalue may be difficult to be proved
experimentally in near future, it seems to deserve further study as an interesting 
possibility, especially, in the scotogenic model.
 
\section*{Appendix~~Jarlskog invariant of the PMNS matrix } 
Although there is no evidence of $CP$ violation in the neutrino oscillation, 
the T2K exteriment \cite{cp-pmns} gives a hint of  $CP$ violation at 2$\sigma$ level 
in the neutrino oscillation. It could result in a larger value for the Jarlskog invariant 
compared to the one obtained in eq.~(\ref{ckm}) \cite{pdg,3nu}.
To respond to the case such a possibility is confirmed, it seems to be useful to consider an extension 
of the present scenario. As such a candidate, we consider an additional mixing between 
the ordinary charged leptons $e_{R_j}$ and heavy vector-like singlet leptons 
$(E_L, E_R)$ such as $F_j\bar E_Le_{R_j}$ \cite{vec}. 
After integrating out the vector-like leptons, the chrged lepton mass matrix 
${\cal M}_e$ gets an additional contribution as
\begin{equation}
{\cal M}_e{\cal M}_e^\dagger - \frac{1}{FF^\dagger+M_E^2}({\cal M}_eF^\dagger)(F{\cal M}_e^\dagger),
\label{mixing}
\end{equation}
where $M_E$ is the mass of the vector-like leptons and satisfies $FF^\dagger >M_E^2$. 
Eq.~(\ref{mixing}) is expected to be diagonalized by $U_e\tilde U$ where $\tilde U$ is a unitary matrix.  
If we consider a case $F=(0, F_1, F_2)$, $\tilde U$ may be roughly approximated as
\begin{equation}
\tilde U=\left(\begin{array}{ccc} 1 & 0 & 0\\
0& \cos\theta & e^{-i\epsilon}\sin\theta\\
0 & -e^{i\epsilon}\sin\theta & \cos\theta \\ 
\end{array}\right), 
\end{equation}
since $U_e$ is supposed to be $V_{\rm CKM}$ which is almost diagonal.
Although this may not be a good approximation, we use it to see a qualitative feature of this kind of 
contribution here.

We examine whether the assumed matrix can reproduce the expected PMNS matrix 
including the Jarlskog invariant. 
If we take $\theta =0.18$ and $\epsilon=-2.0$ as an example, the PMNS matrix is obtained as
\begin{equation}
V=\left(\begin{array}{ccc}0.89 & 0.43 & 0.15\\
0.21 & 0.67 & 0.71\\
0.40 & 0.61 & 0.68 \\ \end{array}\right), 
\label{eckm}
\end{equation}
where all elements represent their absolute values and $J=-0.009$.
The best fit value predicted through the 3$\nu$ global fit is $J=-0.018$,
although there is only small statistical significance between it and $J=0$ \cite{3nu}.
Since $\theta$ and $\epsilon$ are considered to parameterize the coupling $F$,
this scenario could be a promising way to study the origin of the PMNS matrix. 
Both effective masses for the neutrinoless double $\beta$ decay and the $\beta$ decay 
predicted from this result are  
$m_{\beta\beta}= 0.0028$ eV, $m_\beta= 0.0086$ eV in the NO case, and
$m_{\beta\beta}=0.048$ eV,  $m_\beta=0.049$ eV in the IO case.

\newpage

\end{document}